\documentclass[reprint,
amsmath,amssymb,
aps,
]{revtex4-2}
\usepackage{graphicx}
\usepackage{dcolumn}

\begin{document}

\title{
Quantum transport of topological spin solitons\\
in a one-dimensional organic ferroelectric
}

\author{
S. Imajo$^{1,*}$,$\thanks{imajo@issp.u-tokyo.ac.jp}$
A. Miyake$^{1}$,
R. Kurihara$^{1}$,
M. Tokunaga$^{1}$,
K. Kindo$^{1}$,
S. Horiuchi$^{2}$,
and
F. Kagawa$^{3,4}$
}
\affiliation{
$^1$Institute for Solid State Physics, University of Tokyo, Kashiwa, Chiba 277-8581, Japan\\
$^2$Research Institute of Advanced Electronics and Photonics (RIAEP), National Institute of Advanced Industrial Science and Technology (AIST), Tsukuba 305-8565, Japan\\
$^3$RIKEN Center for Emergent Matter Science (CEMS), Wako 351-0198, Japan\\
$^4$Department of Applied Physics, University of Tokyo, Tokyo 113-8656, Japan
}

\date{\today}

\begin{abstract}
We report the dielectric, magnetic, and ultrasonic properties of a one-dimensional organic salt TTF-QBr$_3$I.
These indicate that TTF-QBr$_3$I shows a ferroelectric spin-Peierls (FSP) state in a quantum critical regime.
In the FSP state, coupling of charge, spin, and lattice leads to emergent excitation of spin solitons as topological defects.
Amazingly, the solitons are highly mobile even at low temperatures, although they are normally stationary because of pinning.
Our results suggest that strong quantum fluctuations enhanced near a quantum critical point enable soliton motion governed by athermal relaxation.
This indicates the realization of quantum topological transport at ambient pressure.
\end{abstract}

\maketitle
%  Introduction
 One-dimensional systems exhibit a rich variety of physics related to lattice instabilities through coupling with charge and/or spin degrees of freedom.
 %---------------------
\begin{figure}
\begin{center}
\includegraphics[width=\hsize]{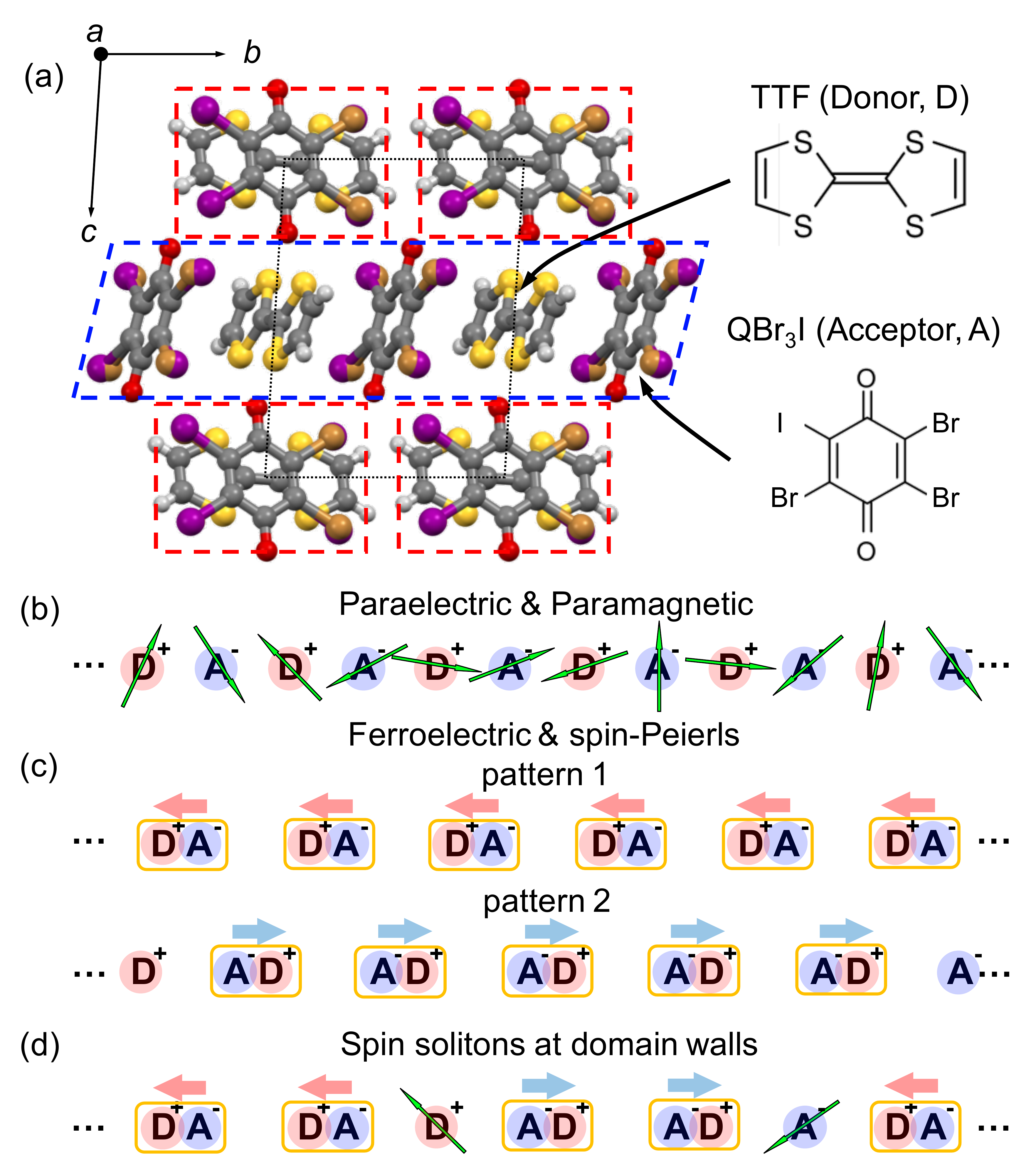}
\end{center}
\caption{
(a) Crystal structure viewed along the $a$-axis, and chemical forms of TTF and QBr$_3$I molecules.
The dashed boxes signify the one-dimensional chains along the $a$-axis (red) and $b$-axis (blue).
(b),(c) Schematic illustrations of the arrangement of the D$^+$ and A$^-$ molecules in (b) the paraelectric paramagnetic state and (c) the FSP state.
The green arrows represent the magnetic spin.
The red and blue arrows signify the directions of the electric dipoles in the D$^+$A$^-$ dimers.
In the FSP state shown in (c), two degenerate patterns occur, patterns 1 and 2, according to the direction of the dipole moments.
(d) Creation of spin solitons at the ferroelectric DWs in the FSP state.
}
\label{fig1}
\end{figure}
%---------------------
The entanglement of multiple degrees of freedom provides intriguing phases and exotic excitations.
One representative example is the spin-Peierls (SP) transition which induces lattice deformation triggered by spin-singlet dimerization.
Whereas this transition has been extensively examined in long-standing theories\cite{1,2,3}, its experimental realization is still limited to only a handful of one-dimensional materials, such as CuGeO$_3$\cite{4}, NaV$_2$O$_5$\cite{5}, and some organic compounds\cite{6,7,8,9}.
Among them, one-dimensional organic charge-transfer complexes have received particular attention because of the strong lattice-charge/spin coupling in molecular crystals.
MEM(TCNQ)$_2$ (MEM=N-methyl-N-ethylmorpholinium, TCNQ=7,7',8,8'-tetracyanoquinodimethane)\cite{6}, TTF-AuS$_4$C$_4$(CF$_3$)$_4$ (TTF=tetrathiafulvalene)\cite{7,8}, (TMTTF)$_2$PF$_6$ (TMTTF=tetramethyltetrathiafulvalene)\cite{9}, $etc$. have been investigated as model systems and have provided significant information on the SP transition, such as the high-field incommensurate phase\cite{8} and pressure-induced quantum criticality\cite{9}.
TTF-QBr$_4$ (QBr$_4$ denotes $p$-bromanil) is also known to undergo the SP transition at 53~K\cite{10,11,12,13}.
However, this salt is quite unique because it is the only example that the SP transition occurs simultaneously with a paraelectric-ferroelectric transition\cite{10,11,12,13,15}.
TTF-QBr$_3$I (2-iodo-3,5,6-tri-bromo-$p$-benzoquinone) focused in this study is isomorphous with TTF-QBr$_4$ although this transition has not been observed\cite{14}.
In these salts, the charge transfer between the donor (D=TTF) and acceptor (A=QBr$_4$ or QBr$_3$I) makes these molecules fully ionic, D$^+$ and A$^-$, in the whole temperature range\cite{10,11,12}.
This means that TTF-QBr$_4$ and TTF-QBr$_3$I are regarded as one-dimensional ionic Mott insulators\cite{15}.
Note that the crystal structure and electronic state of these salts are distinct from those of the other well-known non-magnetic ferroelectrics, TTF-QCl$_4$\cite{21,26,27,32,33}, TTF-QBrCl$_3$\cite{22}, and TTF-QBr$_2$I$_2$\cite{14,28}, which exhibits the neutral-ionic (N-I) transition instead of the SP transition.
This difference manifests in magnetism and electrical conductivity, as discussed in Ref.~\cite{15}

 As displayed in Fig.~\ref{fig1}(a), D$^+$ and A$^-$ are alternately stacked in a one-dimensional chain in TTF-QBr$_4$ and TTF-QBr$_3$I.
At room temperature, the uniform stacking without long-range dimerization provides the paraelectric paramagnetic state (Fig.~\ref{fig1}(b)).
Once the SP transition occurs, the static dimerization alters the paramagnetic state into a non-magnetic state.
The static displacement of D$^+$ and A$^-$ simultaneously leads to ferroelectric order along the chains.
The coupling of the dielectric and magnetic transitions opens up a novel route for magnetic-field-controllable ferroelectrics\cite{12}.
From another viewpoint of the ferroelectric SP (FSP) state, domain formation should be noted because two patterns of opposite dimerization are degenerate, as illustrated by patterns 1 and 2 in Fig.~\ref{fig1}(c).
The two patterns coexist by forming domains, and consequently, domain walls (DWs) are created at their border.
In the case of the N-I ferroelectric systems\cite{14,21,26,27,32,33,22,28}, some excitations, such as a polaron, a N-I DW, a spin soliton, and a charge soliton, have been discussed in terms of topological defects.
On the other hand, in the fully ionic FSP state, only the spin soliton is hosted as the DW as presented in Fig.~\ref{fig1}(d).
This means that we can discuss the pure contribution of the spin soliton, which should be intriguing in terms of topological spin excitation; however, the presence of spin solitons in the FSP state has not yet been observed.
Moreover, the jump of the polarization at the DWs endows the spin solitons with bound charge\cite{39}, and therefore, dynamics of the spin soliton can organize topological transport of spin and charge.
In this work, we examine the dielectric, magnetic, and ultrasonic properties of TTF-QBr$_3$I to discuss the low-temperature emergent phenomena produced by the coupling of charge, spin, and lattice degrees of freedom in a one-dimensional system.
We first discover the FSP state occurring in the quantum critical region.
As expected in the one-dimensional FSP system, the presence of solitonic spins created at the DWs is detected.
Moreover, athermal relaxation between the potential minima of the energy landscape manifests in the low-temperature dynamics due to the strong quantum fluctuations.
These results promise realization of quantum transport of the topological spins in TTF-QBr$_3$I at ambient pressure.

\begin{figure*}
\begin{center}
\includegraphics[width=\hsize]{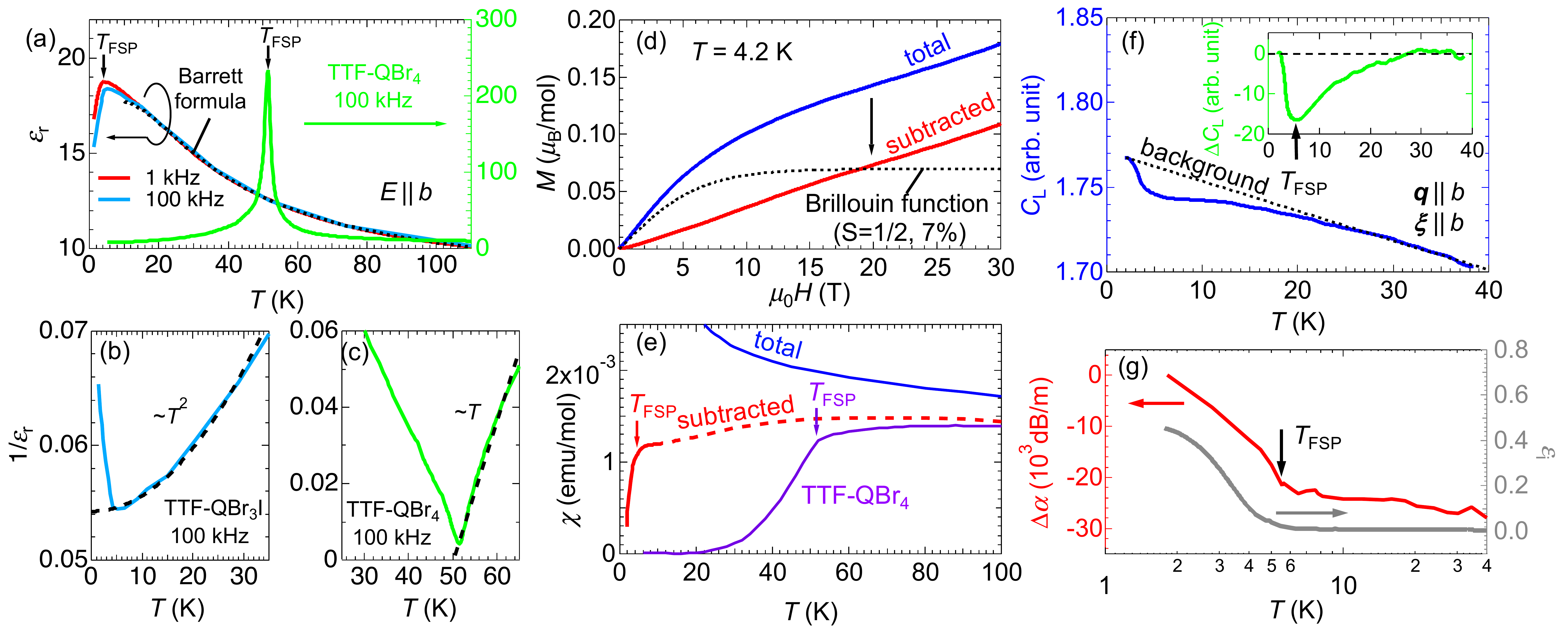}
\end{center}
\caption{
(a) Temperature dependence of the dielectric permittivity in 1~kHz (red) and 100~kHz (light blue) ac electric fields.
The reported data of TTF-QBr$_4$ (light green) are also shown on the right axis.
The dotted curves are the fits to the Barrett formula with the parameters mentioned in the text.
(b),(c) 1/$\epsilon$$_{\rm r}$ vs. $T$ plot of TTF-QBr$_3$I data (b) and TTF-QBr$_4$ data (c) shown in (a).
(d) Magnetization curves up to 30~T at 4.2~K.
The blue curve represents the total magnetization of TTF-QBr$_3$I, while the red curve denotes the magnetization obtained by subtracting the paramagnetic component displayed by the dotted curve.
(e) Total and subtracted magnetic susceptibility as a function of temperature.
The susceptibility is obtained by subtracting the Curie-type paramagnetic component as mentioned in the text and Supplemental materials\cite{suppl}.
Since the subtraction is valid only around 4.2~K, the higher-temperature data is shown as the dashed curve.
The purple curve presents the data for TTF-QBr$_4$.
The arrows signify the FSP transition temperatures.
(f) Temperature dependence of the elastic constant for longitudinal ultrasonic waves along the $b$-axis $C_{\rm L}$.
The dashed curve is a background curve estimated based on the normal elastic stiffening\cite{23}.
The inset displays the additional component related to the FSP transition derived by subtracting the background.
(g) Relative change in temperature-dependent ultrasonic attenuation $\Delta$$\alpha$ plotted in a semilogarithmic plot. 
The data of imaginary part of the dielectric permittivity $\epsilon$$_{\rm i}$ at 1~kHz is also shown on the right axis.
}
\label{fig2}
\end{figure*}

% Result
 First, to discuss the low-temperature state of TTF-QBr$_3$I from the perspective of the dielectric response, we present the temperature dependence of the dielectric permittivity in Fig.~\ref{fig2}(a).
 %---------------------
%---------------------
At 4-5 K (=$T_{\rm FSP}$), the permittivity exhibits an anomaly.
Below 5~K, the permittivity shows the frequency dependence (see Fig.~S1 in Supplemental materials\cite{suppl}), which may arise from the ferroelectric domain dynamics as in the case of other ferroelectrics\cite{14,26,28}.
The frequency-dependent behavior makes the determination of $T_{\rm FSP}$ difficult, but indicates that the macroscopic ferroelectric domains should be formed above 5 K.
The behavior seems to be different from that of typical ferroelectrics such as TTF-QBr$_4$, but, it strongly resembles that of ferroelectricity in the quantum critical regime (quantum ferroelectricity)\cite{14}.
This implies that quantum fluctuations influence the ferroelectricity.
We therefore evaluate the temperature dependence of the permittivity above 10~K by using the Barrett formula for quantum paraelectricity\cite{16}:
\begin{equation}
\epsilon_{\rm r}(T)=C/[(T_{1}/2){\rm coth}(T_{1}/2T)-T_{0}]+A, 
\end{equation}
where $T_0$ and $T_1$ denote the classical Curie-Weiss temperature and the crossover temperature from the classical regime to the quantum-mechanical regime.
The obtained parameters are $T_0$ $\sim$4~K and $T_1$ $\sim$60~K.
The positive value of $T_0$ directly indicates the presence of a ferroelectric interaction.
In addition, the quantum effect on the ferroelectricity is expected to be strong because the ratio between $T_0$ and $T_1$ reaches 15, which is much larger than that of other quantum paraelectrics\cite{17,18}.
To assess whether the ferroelectricity of TTF-QBr$_3$I is in the quantum critical region, the reciprocal permittivity 1/$\epsilon$$_{\rm r}$ is displayed in Fig.~\ref{fig2}(b).
In quantum ferroelectrics, 1/$\epsilon$$_{\rm r}$ varies as $T^2$\cite{14,19,20}, in contrast to the Curie-Weiss behavior 1/$\epsilon$$_{\rm r}$$\sim$$T$ in classical ferroelectrics.
TTF-QBr$_3$I exhibits the quantum critical behavior 1/$\epsilon$$_{\rm r}$$\sim$$T^2$, distinct from the 1/$\epsilon$$_{\rm r}$$\sim$$T$ dependence for classical ferroelectrics such as TTF-QBr$_4$ in Fig.~\ref{fig2}(c).
The dielectric response above 5~K in TTF-QBr$_3$I is governed by the strongly developed quantum fluctuations of the FSP state.
Namely, the chemical substitution from TTF-QBr$_4$ to TTF-QBr$_3$I shifts the ferroelectric transition toward the brink of the quantum critical point (QCP).
Indeed, the low-temperature $\epsilon_{\rm r}$ of TTF-QBr$_3$I is enhanced by the quantum criticality---$\epsilon_{\rm r}$ of TTF-QBr$_3$I becomes twice larger than that of TTF-QBr$_4$ at 5~K.

 Next, to confirm the excitation of spin solitons, we display the magnetization curve at 4.2~K in Fig.~\ref{fig2}(d).
By simply decomposing the $M$-$H$ curve, we obtain the noninteracting paramagnetic component described as the S=1/2 Brillouin function ($\sim$7$\%$) and the almost linear contribution.
The former is considered to originate from spin solitons because the distance between the diluted spin solitons is sufficiently long to disregard the exchange interaction of the solitons\cite{21,22}.
Although it is hard to estimate the number of static impurity spins precisely, the main contribution of the paramagnetic component should be the spin solitons because the number of impurity spins is typically smaller than 1$\%$ in TTF-QX$_4$ salts\cite{12,14,21,22} thanks to the unique molecular shape, which prevents the crystals from having defects and impurities.
The heat capacity measurement also detects the noninteracting component as the two-level-type Schottky anomaly (see Fig.~S2\cite{suppl}), and the value is almost consistent with the value 0.07$\mu$$_{\rm B}$.
We should notice that this value is determined as a static average of the soliton density, which may differ in other time scales depending on the creation and annihilation speed of the solitons.
The almost linear contribution should arise from the antiferromagnetically coupled spins in the TTF and QBr$_3$I chains.
Even if the ferroelectric transition observed at $\sim$5~K is accompanied by the SP transition, the almost linear behavior is reasonable because the transition temperature is quite close to the measurement temperature of 4.2~K.
To clarify that the SP transition simultaneously appears at the same temperature of 4-5~K, in Fig.~\ref{fig2}(e), we present the temperature dependence of the total and subtracted magnetic susceptibility $\chi$ at 1~T after the soliton contribution estimated by the $M$-$H$ curve has been subtracted.
Note that the number of spin solitons should depend on temperature.
Above the FSP transition, the number of spin solitons is smaller, but not zero, because local domain formation exists as a dimerization fluctuation as in the case of TTF-QBr$_4$\cite{12}.
Although the accurate values of $\chi$ are between the subtracted and non-subtracted data, the abrupt decrease of $\chi$ at low temperatures indicates that the transition temperature is almost 5~K.
This behavior evidences the occurrence of the SP transition together with the ferroelectric transition, as in the case of TTF-QBr$_4$.

 Based on the dielectric and magnetic measurement results, we confirm the FSP transition at $\sim$5~K and the presence of spin solitons in TTF-QBr$_3$I.
Since the transition originates from the one-dimensional lattice instability, we next investigate the ultrasonic properties sensitive to lattice deformation.
Figures~\ref{fig2}(f) and 2(g) show the elastic constant $C_{\rm L}$ and the relative change in the ultrasonic attenuation coefficient $\Delta$$\alpha$ for longitudinal ultrasonic waves as a function of temperature, respectively.
$C_{\rm L}$ is known to increases with decreasing temperature due to the normal stiffening of the lattice\cite{23}, regarded as a background component, as denoted by the dashed line in Fig.~\ref{fig2}(f).
Thus, the additional component shown by the green curve in the inset should correspond to the phonon softening due to the FSP transition.
The behavior indicates that fluctuating dimerization grows below 30~K in the high-temperature paramagnetic state as mentioned above and that the long-range dimerization of the SP transition occurs at 5-6 K\cite{24}.
In Fig.~\ref{fig2}(g), the temperature dependence of $\Delta$$\alpha$ also shows an anomaly coming from the FSP transition.
Although $\Delta$$\alpha$ in the SP state usually decreases with decreasing temperature due to the formation of an energy gap\cite{25}, it increases below the transition temperature.
This means that scattering of the acoustic phonons is enhanced in the FSP state.
This behavior makes sense because the emergence of the domain structure increases the scattering rate at the domain boundaries.
Indeed, this temperature dependence is quite similar to that of the imaginary part of the permittivity $\epsilon_{\rm i}$, which reflects energy dissipation by the domain dynamics in ac electric fields, as shown in Fig.~\ref{fig2}(g).
In other words, the scattering between the phonons and spin solitons is promoted with decreasing temperature in the FSP state, as a result of the strong lattice-spin coupling in the present material.

\begin{figure}
\begin{center}
\includegraphics[width=\hsize,clip]{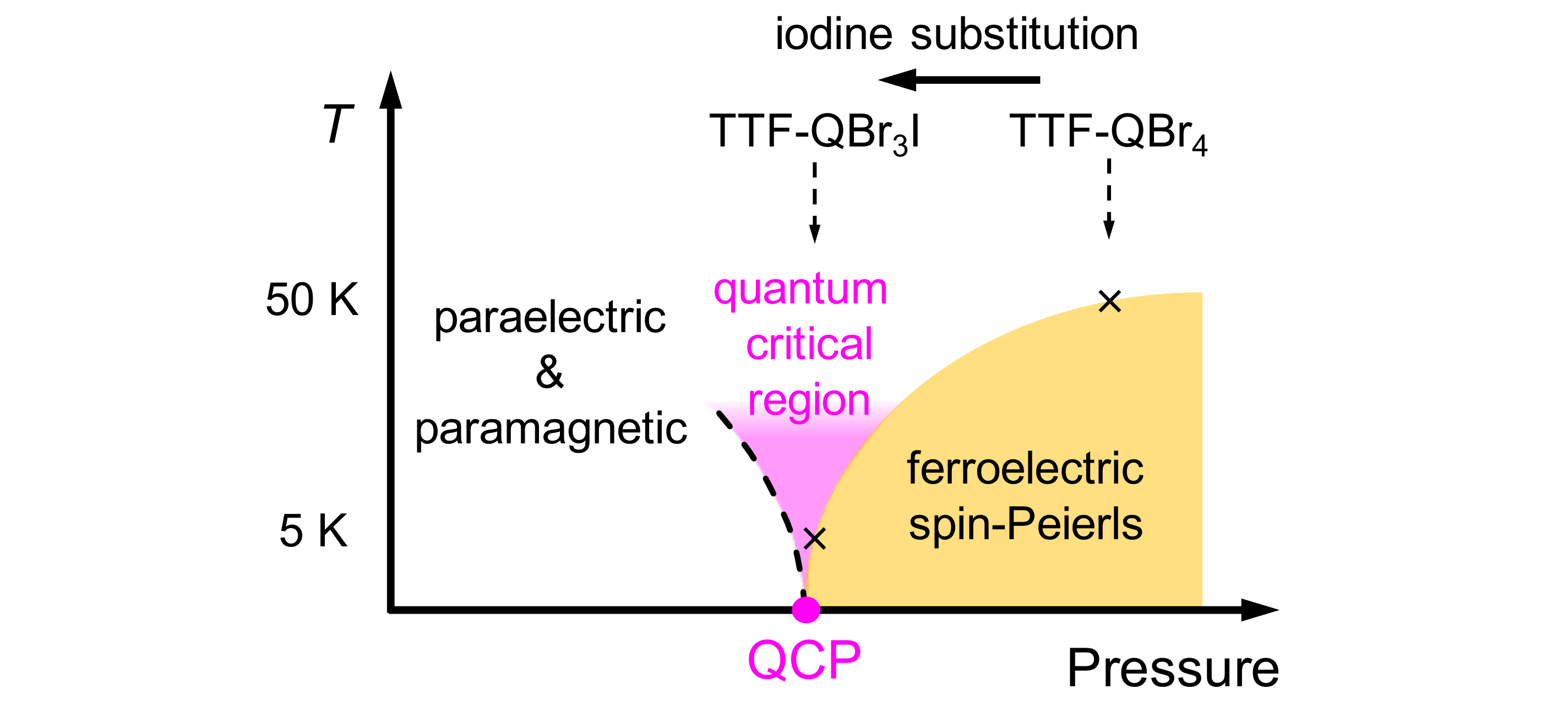}
\end{center}
\caption{
Schematic illustration of the temperature-pressure phase diagram for the FSP system.
TTF-QBr$_3$I is positioned in the quantum critical region located around the QCP.
}
\label{fig3}
\end{figure}

%Discussion
 From our comprehensive investigations, we find that TTF-QBr$_3$I exhibits the FSP transition at $\sim$5~K, clearly detected as dielectric, magnetic, and ultrasonic anomalies.
Interestingly, the low-temperature transition occurs in the quantum critical region, in contrast to the high-temperature FSP transition for TTF-QBr$_4$.
Considering the difference between the two systems, $i.e.$, the halogen atoms Br and I, working as a chemical pressure\cite{14}, the negative chemical pressure thrusts the FSP transition into the quantum critical region, as illustrated in Fig.~\ref{fig3}.
Note that the effect of randomness originating from the replacement with asymmetric molecules on TTF-QX$_4$ is typically less significant than the chemical pressure according to the earlier reports for TTF-QBrCl$_3$\cite{22} and  DMTTF-QBr$_n$Cl$_{4-n}$\cite{25p5}.
This pressure-controllable phase diagram agrees with the typical concept of the quantum criticality for second-order transition between ordered and disordered phases.
The degeneracy of the FSP ground states yields the domain structure, as detected by the augmentation of the ultrasonic attenuation.
In the FSP state the domains produce spin solitons at their boundaries as topological defects.

%---------------------
\begin{figure}
\begin{center}
\includegraphics[width=\hsize,clip]{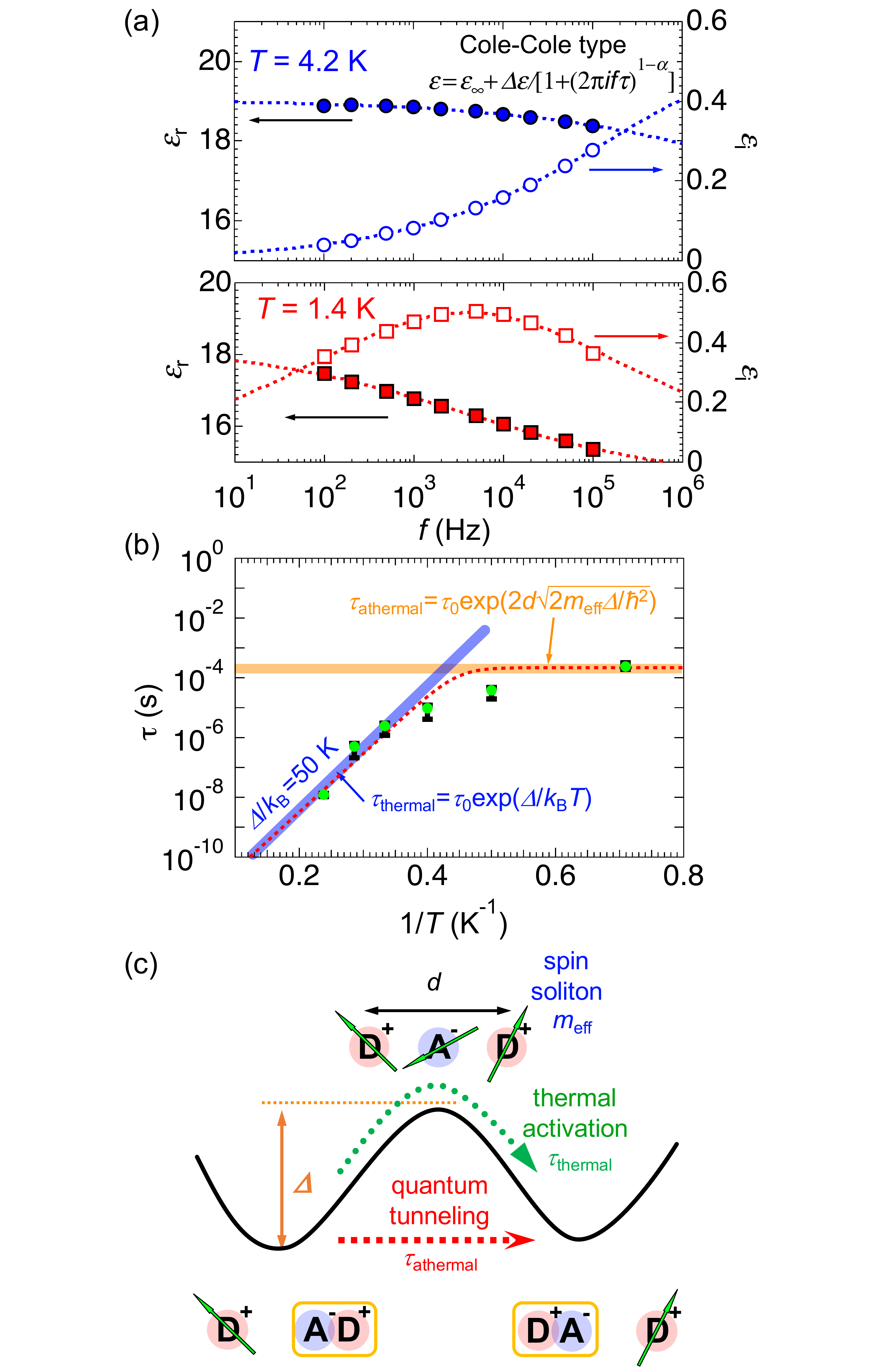}
\end{center}
\caption{
(a) Permittivity-frequency profiles at 1.4~K and 4.2~K.
The dotted curves denote fits to the Cole-Cole type relaxation described in the figure.
(b) Relaxation time as a function of inverse temperature.
The blue line shows the Arrhenius-type linear dependence of the classical relaxation, whereas the orange line is the constant relaxation of the quantum tunneling.
The dotted curve indicates a simple approximation of the crossover between the classical and quantum regime obtained by the Wentzel-Kramers-Brillouin model.
(c) Schematic energy landscape describing the disassociation and recombination of the dimer accompanied by the annihilation and creation of spin solitons.
The parameters d, $\Delta$, and $m_{\rm eff}$ denote the unit cell distance, activation energy and effective mass of the spin solitons, respectively.
$\tau$$_{\rm athermal}$ and $\tau$$_{\rm thermal}$ are the relaxation times of the quantum tunneling process and thermal activation process crossing the potential.
}
\label{fig4}
\end{figure}

%---------------------
 To gain more insight into the FSP state in the quantum critical region, we further scrutinize the low-temperature permittivity in detail below.
As shown in Fig.~\ref{fig2}(a), the permittivity exhibits a frequency dependence at low temperatures.
This behavior arises from the dynamics of the ferroelectric domains similar to those in other organic ferroelectrics\cite{26,27,28}.
This means that the frequency dependence induced by the DW dynamics directly reflects the soliton motion.
The characteristic relaxation time $\tau$ can be derived by examining at the frequency dependence of the dielectric permittivity shown in Fig.~\ref{fig4}(a).
%---------------------
%---------------------
The behavior is well reproduced by one mode of the Cole-Cole-type relaxation shown in the figure\cite{29}, and from this analysis, we obtain $\tau$ as a function of inverse temperature shown in Fig.~\ref{fig4}(b).
The stretching parameter in the relaxation equation, $\alpha$, is $\sim$0.7 in this temperature region.
The large value of $\alpha$ is consistent with the enhancement of $\alpha$ with approaching the ferroelectric QCP observed in other quantum ferroelectrics\cite{28}.
This means that the developed quantum fluctuations make the spectral width of the DW response broad.
To shift a DW, recombination of the dimer is required, as is illustrated in Fig.~\ref{fig4}(c).
The energy for the dimer dissociation corresponds to the activation energy $\Delta$, which acts as an effective pinning mechanism and results in the thermal activation behavior of the soliton motion.
Thus, the relaxation time of the domain dynamics is exponentially suppressed with decreasing temperature.
The linear dependence of $\tau$ below $\sim$0.4~K$^{-1}$ (above $\sim$2.5~K) in this plot exactly demonstrates the Arrhenius-type behavior of the dynamics, indicating slowing of the soliton motion towards low temperatures.
However, surprisingly, the decrease in $\tau$ deviates from the linear dependence at low temperatures, 1/$T$$>$$\sim$0.4~K$^{-1}$ ($i.e.$, $T$$<$$\sim$2.5~K), and the fast relaxation ($\tau$ $\sim$10$^{-4}$~s) seems to survive even in the zero-temperature limit.
This means that the spin solitons are highly mobile without suffering from pinning even at low temperatures, which is in marked contrast to the typical dynamic freezing of glasses described by the Vogel-Fulcher-Tammann equation\cite{29p5}.
Similar behavior has been reported in the previous work on the quantum ferroelectric state of the ferroelectric N-I transition\cite{28}.
Those researchers concluded that the ferroelectric DWs creep in an athermal process dominated by the quantum fluctuations enhanced near the QCP.
Although the magnetic degree of freedom is quenched in the ferroelectric N-I transition because of the simultaneous charge transfer, the similar response indicates that the spin solitons in TTF-QBr$_3$I are also transmitted across the potential landscape by quantum tunneling.
Thus, in the same manner, we evaluate the dynamics of the spin solitons with a simple model, the Wentzel-Kramers-Brillouin approximation\cite{28,30} for quantum tunneling and Matthiessen's rule by the following formula:
\begin{equation}
\begin{split}
\tau(T)=(1/\tau_{\rm thermal}+1/\tau_{\rm athermal})^{-1}\\
\quad=\tau_{0}/[{\rm exp}(-\Delta /k_{\rm B}T)+{\rm exp}(-2d\sqrt{2m_{\rm eff}\Delta / \hbar^{2}})],
\end{split}
\end{equation}
where $d$ signifies the tunneling distance of the soliton, namely, the unit cell length along the column, $\approx$8.5~{\AA}\cite{14}, and $m_{\rm eff}$ represents the effective mass of the spin soliton.
$\tau$$_0$ is the attempt relaxation time.
The first term $\tau$$_{\rm thermal}$ represents the relaxation time of the Arrhenius-type relaxation, while the second term $\tau$$_{\rm athermal}$ denotes that of the quantum relaxation.
The behavior cannot be completely described by the present simple approximation depicted by the red dotted curve, but the assumption roughly gives some parameters related to the dynamics.
The estimated values of $m_{\rm eff}$ and $\Delta$/$k_{\rm B}$ are $\sim$1000$m_{\rm e}$ ($m_{\rm e}$ is the electron mass) and $\sim$50~K, respectively.
These two are the origin of the fast $\tau$$_{\rm athermal}$.
Since the tunneling of the spin solitons involves displacement of the molecules, $m_{\rm eff}$ should be on the order of the masses of TTF and QBr$_3$I (10$^5$$m_{\rm e}$-10$^6$$m_{\rm e}$); however, the obtained $m_{\rm eff}$ is several hundred times smaller than the expected value.
In earlier reports on the soliton/DW dynamics\cite{28,31}, a similar drastic diminishment was observed and discussed from the viewpoint of the soliton width.
The decrease in $\Delta$ when approaching the QCP causes broadening of the DW width with the development of quantum fluctuations.
For TTF-QBr$_4$, the previous work\cite{12} reported a spin gap value of $\Delta$/$k_{\rm B}$$\sim$250~K, which should be comparable with $\Delta$ because both the gaps are the energy difference between the order and disorder states.
The approach to the QCP certainly gives the much smaller $\Delta$/$k_{\rm B}$$\sim$50~K for TTF-QBr$_3$I, which reasonably reduces the effective mass by the strong broadening of the soliton width.
Accordingly, the nearness to the QCP, giving the light $m_{\rm eff}$ and small $\Delta$, entails the fast dynamics of the spin solitons, indicative of the quantum transport of the topological spin solitons.

% Conclusion
The present results substantiate that the FSP state of TTF-QBr$_3$I is inside the quantum critical region.
The topological spin solitons in the FSP state are endowed with high mobility even in the low-temperature region owing to the strengthened quantum fluctuations.
The pure transport of the spin solitons induced by the quantum fluctuations must materialize in TTF-QBr$_3$I.
This quantum transport is distinct from the DWs thermally traveling near a room-temperature critical point in the N-I ferroelectric salt TTF-QCl$_4$\cite{27,32,33}.
Since this promises unique transport mediated by the flowing spin solitons, further studies, such as thermal transport measurements, are the interesting subjects for future work.

We thank Y. Nemoto and M. Akatsu (Niigata University) for supplying the LiNbO$_3$ piezoelectric transducers used in this study.
This study was partly supported by JST CREST Grant Number JPMJCR18J2.

\renewcommand{\thefigure}{S\arabic{figure}}
\clearpage
\onecolumngrid
\appendix
\begin{center}
\large{\bf{Supplemental Materials for\\
Quantum transport of topological spin solitons\\
in a one-dimensional organic ferroelectric
}}
\end{center}
\section{Experimental methods}
Single crystals of TTF-QBr$_3$I were grown by slow evaporation of cold ($\sim$5~$^{\circ}$C) mixed acetonitrile solution of QBr$_3$I and TTF and harvested as the minor products with the main products of (TTF)$_2$QBr$_3$I.
While TTF-QBr$_3$I crystallized as black rectangular plates, (TTF)$_2$QBr$_3$I crystallized as dark brown blocks.
The dielectric permittivity was measured by means of the typical two-terminal method along the b-axis parallel to the one-dimensional columns in the frequency range of 100~Hz-100~kHz.
The magnetization measurement for the $M$-$H$ curve was carried out in a pulse magnet with polycrystalline samples (total weight:$\sim$30~mg).
We performed the static magnetic susceptibility measurement at 1~T by using a magnetic property measurement system (Quantum Design) with polycrystalline samples weighing about 30~mg.
By using the pulse-echo and the phase comparison methods, the ultrasonic properties were measured. 32.0~MHz longitudinal ultrasonic waves along the $b$-axis are generated and detected by LiNbO$_3$ piezoelectric transducers glued on the (0~1~0) surfaces.

\section{Frequency and temperature dependence of $\epsilon$ }
%---------------------
\begin{figure}[h]
\begin{center}
\includegraphics[width=\hsize,clip]{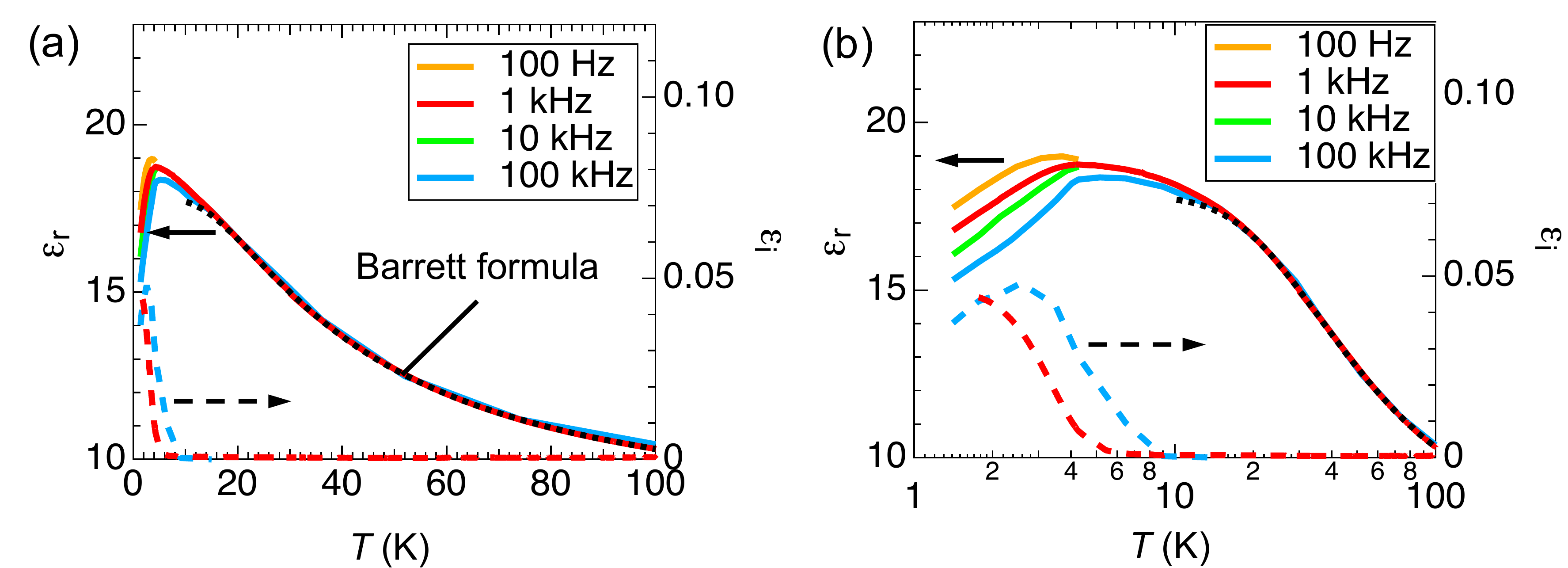}
\end{center}
\caption{Temperature dependence of $\epsilon_{\rm r}$ (left axis) and $\epsilon_{\rm i}$ (right axis) at each frequency in a linear plot (a) and in a semi-logarithmic plot (b).
}
\label{figS1}
\end{figure}
%---------------------
\clearpage

\section{Heat capacity}
%---------------------
\begin{figure}[b]
\begin{center}
\includegraphics[width=0.7\linewidth,clip]{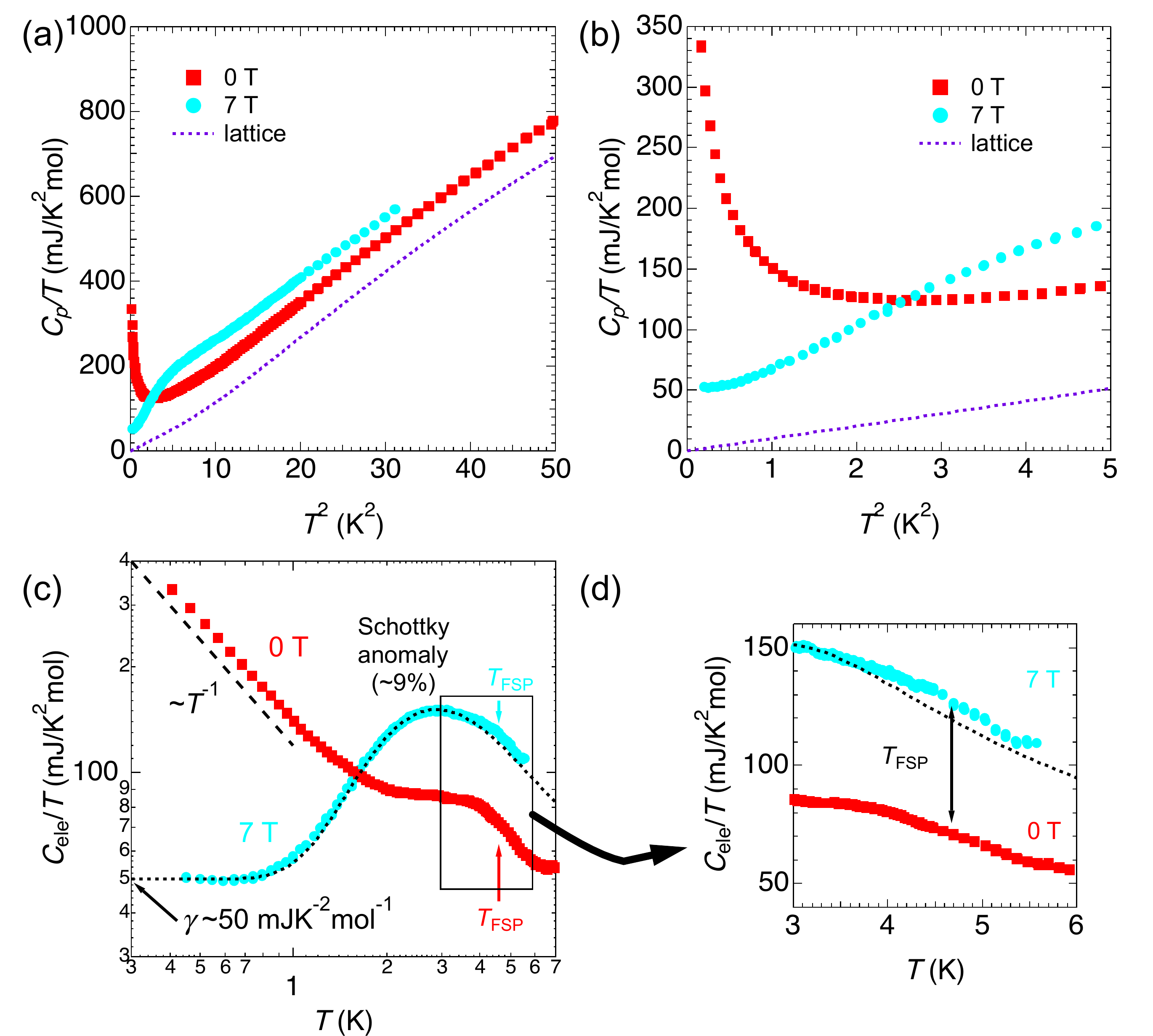}
\end{center}
\caption{
(a)The heat capacity data in the $C_p$/$T$ vs. $T^2$ plot at 0~T and 7~T.
The purple dotted curve denotes the lattice heat capacity mentioned in the text.
(b) The enlarged plot of (a) in the low temperature region.
(c) The electronic part of the heat capacity in a logarithmic plot of  $C_{\rm ele}$/$T$ vs. $T$.
The black dotted curve indicates a fit composed of the constant value $\gamma$ and the two-level Schottky anomaly at 7~T.
The black dashed line is a guide showing the relation  $C_{\rm ele}$/$T$$\sim$$T^{-1}$.
(d) The enlarged plot of (c) around $T_{\rm FSP}$.
}
\label{figS2}
\end{figure}
%---------------------
 Employing a physical property measurement system (Quantum Design), the heat capacity was measured with several single crystals whose total weight is about 5~mg.
We show the total heat capacity plotted as $C_p$/$T$ vs. $T^2$ in Figs.~\ref{figS2}(a),(b).
In the case of the typical organic conductors, the low-temperature heat capacity is described by the following formula:
 \begin{equation}
C_p/T=\gamma+\beta T^{2}+cR(T_{\rm E}/T)^{2}{\rm exp}(T_{\rm E}/T)/T[{\rm exp}(T_{\rm E}/T)-1]^{2}, 
\end{equation}
where $\gamma$ and $\beta$ denote the electronic and lattice heat capacity coefficients, respectively.
The third term, the Einstein mode, represents the optical phonon contribution, coming from the libration motion of the planar molecules\cite{34} like TTF and QBr$_3$I in the present case.
This component is typically negligible at low temperatures below 5~K$^2$.
For the present case, we need to take account of the additional term originating from the spin solitons, which shows the two-level Schottky anomaly $C_{\rm Sch}$ in magnetic fields.
A best fit for the data at 7~T below 5~K$^2$ is given by the parameters, 49.8~mJK$^{-2}$mol$^{-1}$ for $\gamma$, 10.2~mJK$^{-4}$mol$^{-1}$ for $\beta$ and 29.4~K for $T_{\rm E}$.
The Schottky heat capacity yields the density of the spin solitons of $\sim$9$\%$, consistent with the value observed in the magnetization measurement.
Note, however, that there are many parameters to fit the data and we assumed that the contribution of the transition is smaller than these components.
The density of the spin solitons may be overestimated because of the anomalous heat capacity of the transition.
The values of $\beta$ and TE are comparable with those of typical organic charge-transfer salts\cite{34,35}.
To obtain the electronic contributions, the lattice contributions (the dashed curve displayed in Figs.~\ref{figS2}(a),(b)), the second and third terms in the formula, are subtracted from the total heat capacity.
Figure~\ref{figS2}(c) shows the temperature profile of the obtained electronic heat capacity $C_{\rm ele}$ in the logarithmic plot of  $C_{\rm ele}$/$T$ vs. $T$.
At 0~T, the broad hump at 4-5~K is observed.
As identified in the main text, the thermodynamic anomaly reflects the ferroelectric SP transition.
Applying a magnetic field of 7~T, the anomaly becomes hard to distinguish due to the emergence of the large Schottky anomaly.
Nevertheless, the size of the anomaly does not show large field dependence, as shown in Figs.~\ref{figS2}(d).
Apart from the anomaly, we should notice the change of the low-energy excitations.
At low temperatures,  $C_{\rm ele}$/T varies as $\sim$$T^{-1}$ at 0~T while it is constant value $\gamma$ at 7~T.
The heat capacity is expected to exhibit gapped behavior\cite{36} because the ground state is the SP state.
However, in the present case the transition can be no longer treated in the mean-field due to the strong quantum fluctuation.
That's why the thermodynamic anomaly is strongly broadened or smeared.
Since TTF-QBr$_3$I is composed of the one-dimensional antiferromagnetic chains, the Nambu-Goldstone mode has the relation $C_p$$\sim$$T$ in the low temperature region, which is consistent with the constant behavior of  $C_{\rm ele}$/$T$ at 7~T.
Since the coefficient $\gamma$ is scaled with the inverse of $J$/$k_{\rm B}$ in such case, the obtained $\gamma$ give when $J$/$k_{\rm B}$$\sim$200~K, which is roughly comparable to the value determined in the susceptibility measurement.
In any case, the observed $\gamma$ at 7~T can be attributed to the low-energy excitation of the one-dimensional antiferromagnetism.
By contrast, the origin of the diverging heat capacity  $C_{\rm ele}$/$T$$\sim$$T^{-1}$ in zero field is still unclear because the Schottky heat capacity does not appear at zero field.
The disappearance of the component at 7~T indicates that it should be related to magnetic degrees of freedom, such as the spin soliton, the one-dimensional antiferromagnetism and nuclear spin.
Even though internal or remnant fields exist in the sample, the typical Schottky heat capacity exhibits  $C_{\rm ele}$/$T$$\sim$$T^{-3}$ dependence.
Thus, characteristic behavior originating from temperature-dependent relaxation time of the spin solitons and/or nuclear spins or critical behavior of magnetic degrees of freedom are considered as possible candidates.
We should again notice that the present fits have large ambiguity, and therefore, the evaluation of these components should not be accurate quantitatively. 
Although decomposing the total heat capacity into each part is quite hard, the anomalous heat capacity must contain all the degrees of freedom including the contribution of the electric dipoles forming the ferroelectricity.

\section{Raw magnetic susceptibility}
%---------------------
\begin{figure}[hh]
\begin{center}
\includegraphics[width=0.7\linewidth,clip]{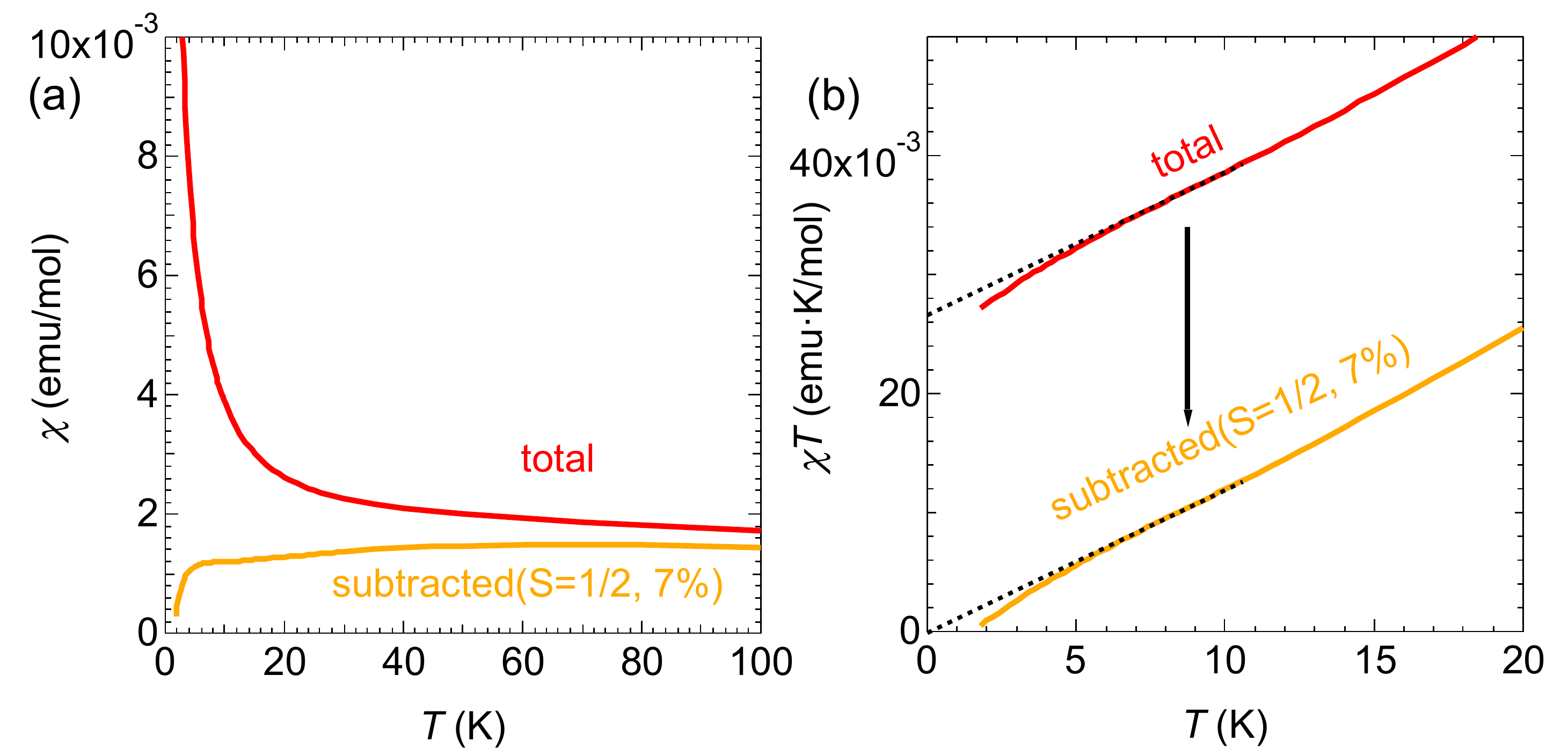}
\end{center}
\caption{
Temperature dependence of magnetic susceptibility of TTF-QBr$_3$I in the plot (a) $\chi$ vs. $T$ and (b) $\chi$$T$ vs. $T$.
The red curve presents the raw data while the orange denotes the data plotted in Fig. 2(b).
The dotted lines in (b) are extrapolation of the data in the range of 7-10~K to 0~K.
}
\label{figS3}
\end{figure}
%---------------------
 As discussed in the main text and shown in Fig.~2(d), TTF-QBr$_3$I has the paramagnetic component of the spin solitons described by the S=1/2 Brillouin function.
While Fig.~2(e) presents the extracted magnetic susceptibility to emphasize the transition, Fig.~\ref{figS3} shows the total magnetic susceptibility (red) without the subtraction of the Curie-type paramagnetic component with the subtracted data.
We should notice that the number of the spin solitons should depend on temperature.
The temperature dependence of the subtracted susceptibility should be improper at higher temperatures far from 4.2~K.
However, the pulsed-field magnetization measurements of the present salt were difficult above 4.2~K, namely the liquid $^4$He temperature, because the small sample signal was undetectable precisely without a highly stable heat bath.
Above the FSP transition, the number of spin solitons is smaller, but not zero, because local domain formation exists as a dimerization fluctuation as evidenced by the elastic data and dielectric data.
This means that the accurate values of $\chi$ above $T_{\rm FSP}$ is between the total $\chi$ data and subtracted $\chi$ data.
Nevertheless, we can evaluate whether the present subtraction is reasonable or not.
By employing the typical manner for one-dimensional systems, the temperature dependence above the FSP transition should be described by the uniform one-dimensional antiferromagnetic Heisenberg model (1D-AFHM).
The absolute value of the subtracted $\chi$ at 4.2~K, which is almost constant below 10~K, is about 1.1~emu/mol.
This value gives the exchange interaction as $J$/$k_{\rm B}$$\sim$150 K\cite{37,38}, roughly consistent with the value 200~K estimated by the linear term of heat capacity even though the analyses of $C_p$ have some fitting ambiguity.
Therefore, the abrupt decrease of $\chi$ below 4.2 K should be reasonable although the accurate values above $T_{\rm FSP}$ are unclear.

\section{Ultrasonic echoes}
\begin{figure}[hh]
\begin{center}
\includegraphics[width=0.4\linewidth,clip]{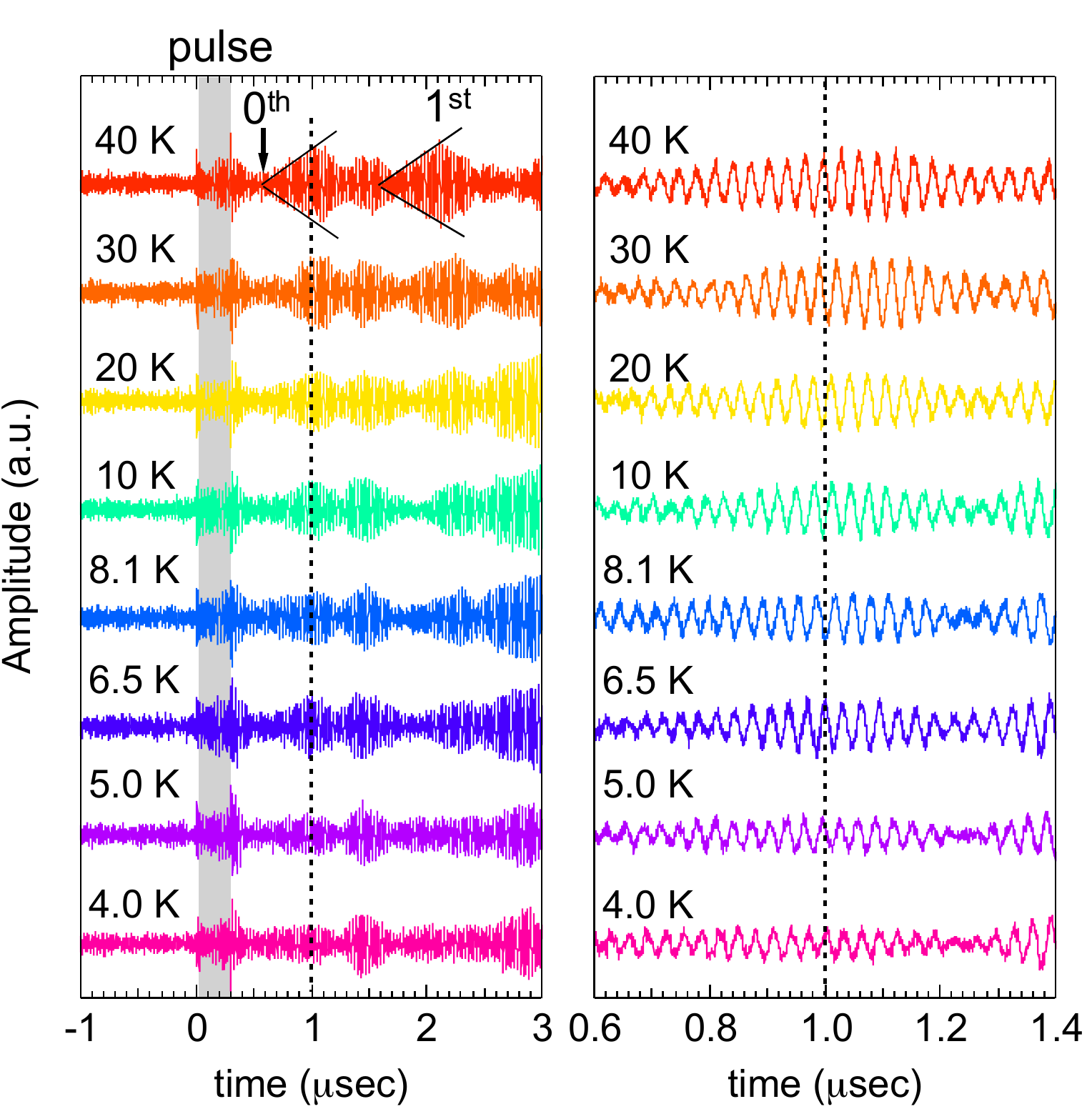}
\end{center}
\caption{
In this measurement, we apply a pulsed ultrasound in the gray area.
The solid lines are guides to estimate the beginning position of the echoes.
The dashed line indicates the position that we analyze the ultrasonic properties of the 0th echo.
The right panel is the enlarged plot of the left panel at 1~$\mu$sec.
}
\label{figS4}
\end{figure}

 In this study, we determine the elastic constant $C$ and the relative change of the ultrasonic attenuation $\Delta$$\alpha$ from the detected ultrasonic echoes.
Figure~\ref{figS4} shows the ultrasonic echoes at various temperatures. In each echo measurement, we give a 300~nsec pulse of 32~MHz to one of the attached transducers at 0~$\mu$sec as described by the shaded area.
The applied ultrasound passes through the sample and reach the opposite side of the sample, which has another attached transducer to detect the echoes as presented in the Fig.~\ref{figS4}.
Thus, we derived the absolute value of the sound velocity $V$ from the sample size and the time span to deliver the 0th echo as 1600$\pm$200~m/sec.
The elastic constant $C$ is given as $\rho$$V^2$, where $\rho$ denotes the density.
For $\Delta$$\alpha$, we evaluate the change of the echo amplitude from that at the lowest temperature.
In order to precisely determine $V$ and $\Delta$$\alpha$, we measured the change of the amplitude and phase of the 0th echo at 1~$\mu$sec presented by the dashed line because the other echoes may include additional components coming from other echoes passing through other routes.
%---------------------
%---------------------


\begin{thebibliography}{99}
%%%
\bibitem{1} M. C. Cross and D. S. Fisher, Phys. Rev. B {\bf19}, 402 (1979).
\bibitem{2} M. C. Cross, Phys. Rev. B {\bf20}, 4606 (1979).
\bibitem{3} J. W. Bray, L. V. Interrante, I. S. Jacobs, and J. C. Bonner, Springer, Boston, MA (1983).
\bibitem{4} M. Hase, I. Terasaki, and K. Uchinokura, Phys. Rev. B {\bf70}, 3651 (1993).
\bibitem{5} M. Isobe and Y. Ueda, J. Phys. Soc. Jpn. {\bf65}, 1178 (1996).
\bibitem{6} S. Huizinga, J. Kommandeur, G. A. Sawatzky, B. T. Thole, K. Kopinga, W. J. M. de Jonge, and J. Roos, Phys. Rev. B {\bf19}, 4723 (1979).
\bibitem{7} I. S. Jacobs, J. W. Bray, H. R. Hart, Jr., L. V. Interrante, J. S. Kasper, G. D. Watkins, D. E. Prober, and J. C. Bonner, Phys. Rev. B {\bf14}, 3036 (1976).
\bibitem{8} J. A. Northby, H. A. Groenendijk, L. J. de Jongh, J. C. Bonner, I. S. Jacobs, and L. V. Interrante, Phys. Rev. B {\bf25}, 3215 (1982).
\bibitem{9} D. S. Chow, P. Wzietek, D. Fogliatti, B. Alavi, D. J. Tantillo, C. A. Merlic, and S. E. Brown, Phys. Rev. Lett. {\bf81}, 3984 (1998).
\bibitem{10} A. Girlando, C. Pecile, and J. B. Torrance, Solid State Commun. {\bf54}, 753 (1985).
\bibitem{11} Y. Tokura, S. Koshihara, Y. Iwasa, H. Okamoto, T. Komatsu, T. Koda, N. Iwasawa, and G. Saito, Phys. Rev. Lett. {\bf63}, 2405 (1989).
\bibitem{12} F. Kagawa, S. Horiuchi, M. Tokunaga, J. Fujioka, and Y. Tokura, Nat. Phys. {\bf6}, 169 (2010).
\bibitem{13} S. Horiuchi, K. Kobayashi, R. Kumai, and S. Ishibashi, Chem. Lett. {\bf43}, 26-35 (2014).
\bibitem{15} K. Sunami, Y. Sakai, R. Takehara, H. Adachi, K. Miyagawa, S. Horiuchi, and K. Kanoda, Phys. Rev. Res. {\bf2}, 043333 (2020). 
\bibitem{14} S. Horiuchi, K. Kobayashi, R. Kumai, N. Minami, F. Kagawa, and Y. Tokura, Nat. Commun. {\bf6}, 7469 (2015).
\bibitem{21} T. Mitani, G. Saito, Y. Tokura, and T. Koda, Phys. Rev. Lett. {\bf53}, 842 (1984).
\bibitem{26} H. Okamoto, T. Mitani, Y. Tokura, S. Koshihara, T. Komatsu, Y. Iwasa, T. Koda, and G. Saito, Phys. Rev. B {\bf43}, 8224 (1991).
\bibitem{27} K. Sunami, T. Nishikawa, K. Miyagawa, S. Horiuchi, R. Kato, T. Miyamoto, H. Okamoto, and K. Kanoda, Sci. Adv. {\bf4}, eaau7725 (2018).
\bibitem{32} R. Takehara, K. Sunami, K. Miyagawa, T. Miyamoto, H. Okamoto, S. Horiuchi, R. Kato, and K. Kanoda, Sci. Adv. {\bf5}, eaax8720 (2019).
\bibitem{33} R. Takehara, K. Sunami, F. Iwase, M. Hosoda, K. Miyagawa, T. Miyamoto, H. Okamoto, and K. Kanoda, Phys. Rev. B {\bf98}, 054103 (2018).
\bibitem{22} F. Kagawa, S. Horiuchi, H. Matsui, R. Kumai, Y. Onose, T. Hasegawa, and Y. Tokura, Phys. Rev. Lett. {\bf104}, 227602 (2010).
\bibitem{28} F. Kagawa, N. Minami, S. Horiuchi, and Y. Tokura, Nat. Commun. {\bf7}, 10675 (2016).
\bibitem{39} P. S. Bednyakov, T. Sluka, A. K. Tagantsev, D. Damjanovic, and N. Setter, Sci. Rep. {\bf5}, 15819 (2015).
\bibitem{suppl} See Supplemental Materials for the detailed analyses of the present measurements and additional information, which includes Refs. [25-29].
\bibitem{34} J. Wosnitza, X. Liu, D. Schweitzer, and H. J. Keller, Phys. Rev. B {\bf50}, 12747 (1994).
\bibitem{35} S. Imajo, N. Kanda, S. Yamashita, H. Akutsu, Y. Nakazawa, H. Kumagai, T. Kobayashi, and A. Kawamoto, J. Phys. Soc. Jpn. {\bf85}, 043705 (2016).
\bibitem{36} W. H. Korving, G. J. Kramer, R. A. Steeman, H. B. Brom, L. J. De Jongh, M. Fujita, and K. Machida, Physica B+C {\bf145}, 299 (1987).
\bibitem{37} J. C. Bonner and M. E. Fisher, Phys. Rev. {\bf135}, A640 (1964).
\bibitem{38} R. B. Griffiths, Phys. Rev. {\bf133}, A768 (1964).
\bibitem{23} Y. P. Varshni, Phys. Rev. B {\bf2}, 3952 (1970).
\bibitem{16} J. H. Barrett, Phys. Rev. {\bf86}, 118 (1952).
\bibitem{17} K. A. M$\ddot{\rm u}$ller and H. Burkard, Phys. Rev. B {\bf19}, 3593(1979).
\bibitem{18} M. Shimozawa, K. Hashimoto, A. Ueda, Y. Suzuki, K. Sugii, S. Yamada, Y. Imai, R. Kobayashi, K. Itoh, S. Iguchi, M. Naka, S. Ishihara, H. Mori, T. Sasaki, and M. Yamashita, Nat. Commun. {\bf8}, 1821 (2017).
\bibitem{19} N. Das and S. G. Mishra, J. Phys. Conden. Mater. {\bf21}, 095901 (2009).
\bibitem{20} S. Rowley, L. J. Spalek, R. P. Smith, M. P. M. Dean, M. Itoh, J. F. Scott, G. G. Lonzarich, and S. S. Saxena, Nat. Phys. {\bf10}, 367 (2014).
\bibitem{24} The relatively higher transition temperature detected here is due to the higher measurement frequency (32~MHz) compared to the frequency for other measurements. The frequency dependence is evident in the permittivity measurement.
\bibitem{25} M. Poirier, M. Castonguay, A. Revcolevschi, and G. Dhalenne, Phys. Rev. B {\bf52}, 16058 (1995).
\bibitem{25p5} S. Horiuchi, Y. Okimoto, R. Kumai, and Y. Tokura, Science {\bf299}, 229 (2003).
\bibitem{29} K. S. Cole and R. H. Cole, J. Chem. Phys. {\bf9}, 341 (1941).
\bibitem{29p5} P. Lunkenheimer, S. Kastner, M. K$\ddot{\rm O}$hler, and A. Loidl, Phys. Rev. E {\bf81}, 051504 (2010).
\bibitem{30} J. Brooke, T. F. Rosenbaum, and G. Aeppli, Nature {\bf413}, 610 (2001).
\bibitem{31} B. Champagne, E. Deumens, and Y. $\ddot{\rm O}$hrn, J. Chem. Phys. {\bf107}, 5433 (1997).
\end{thebibliography}
\end{document}